\documentclass[conference]{IEEEtran}
\IEEEoverridecommandlockouts
\usepackage{cite}
\usepackage{float}
\usepackage{amsmath,amssymb,amsfonts}
\usepackage{algorithmic}
\usepackage{graphicx}
\usepackage{textcomp}
\usepackage{xcolor}
\usepackage{array}
\usepackage{makecell}
\usepackage{graphicx}
\usepackage{multirow}
\usepackage{comment}
\usepackage{threeparttable}
\def\BibTeX{{\rm B\kern-.05em{\sc i\kern-.025em b}\kern-.08em
		T\kern-.1667em\lower.7ex\hbox{E}\kern-.125emX}}

\newcommand{\bfy}{\mathbf{y}}
\newcommand{\bfx}{\mathbf{x}}

\newcommand{\bfu}{\mathbf{u}} 
\newcommand{\bfU}{\mathbf{U}}

\newcommand{\mbf}[1]{\mathbf{#1}}

\newcommand{\what}{\widehat}
\newcommand{\wtilde}{\widetilde}

\DeclareMathOperator*{\argmin}{arg\,min}

\newcommand{\transpose}{\mathsf{T}}

\newcommand{\bfzeros}{\mathbf{0}}
\newcommand{\vertiii}[1]{{\left\vert\kern-0.25ex\left\vert\kern-0.25ex\left\vert #1 
		\right\vert\kern-0.25ex\right\vert\kern-0.25ex\right\vert}}

\newcommand{\bfA}{\mathbf{A}}
\newcommand{\bfB}{\mathbf{B}}

\newcommand{\bfC}{\mathbf{C}}

\newcommand{\norm}[1]{\left\lVert#1\right\rVert}

\newtheorem{theorem}{\bf \emph{Theorem}}

\newtheorem{assumption}[theorem]{Assumption}

\usepackage[linesnumbered,ruled,vlined]{algorithm2e}

\begin{document}
	
	\title{A Complex-LASSO Approach for Localizing Forced Oscillations in Power Systems \\
		\thanks{This material is based upon work supported by the National Science Foundation  under  Grant  No.  OAC-1934766 and PSERC project S-87.}
	}
	
	\author{\IEEEauthorblockN{Rajasekhar Anguluri, Nima~Taghipourbazargani, Oliver Kosut, and Lalitha Sankar}
		\IEEEauthorblockA{\textit{School of Electrical, Computer, and Energy
				Engineering} \\
			\textit{Arizona State University, Tempe, AZ 85281 USA}\\
			\{rangulur,ntaghip1,okosut,lalithasankar\}@asu.edu}}
	
	\maketitle
	
	\begin{abstract}
		We study the problem of localizing multiple sources of forced oscillations (FOs) and estimating their characteristics, such as frequency, phase, and amplitude, using noisy PMU measurements. For each source location, we model the input oscillation as a sum of unknown sinusoidal terms. This allows us to obtain a linear relationship between measurements and the inputs at the unknown sinusoids' frequencies in the frequency domain. We determine these frequencies by thresholding the empirical spectrum of the noisy measurements. Assuming sparsity in the number of FOs' locations and the number of sinusoids at each location, we cast the location recovery problem as an $\ell_1$-regularized least squares problem in the complex domain---i.e., complex-LASSO (linear shrinkage and selection operator). We numerically solve this optimization problem using the complex-valued coordinate descent method, and show its efficiency on the IEEE 68-bus, 16 machine and WECC 179-bus, 29-machine systems. 
	\end{abstract}
	
	
	\section{Introduction}
	Detecting and localizing forced oscillations (FOs) is crucial for safety and reliability of power systems. Detection helps determine if the manifested oscillations in PMU measurements are FOs; whereas localization helps to mitigate the oscillations either by disabling the sources (e.g., malfunctioned controllers and cyclic loads) or by injecting certain counteracting signals\cite{MG:17, DJT-RG:20}. Compared to localization, detecting oscillations is easier and has been well studied in power systems \cite{SR-WJ-NN-BL:21, HY-YL-PZ-ZD:17}. 
	
	We focus on localizing the sources of FO, which amounts to finding $m^*$ true sources from $m$ possible sources. A brute-force search requires searching across ${m \choose m^*}$ configurations, which is computationally intractable for large $m$. One way to tackle this challenge is to find source locations by minimizing performance measures, such as system-theoretic norms (e.g., $\mathcal{H}_2$ and $\mathcal{H}_\infty$) and information-theoretic based measures \cite{KM-JNK-SLB:21}.
	
	Our approach aims to leverage the fact that the FO sources are sparse relative to the number of possible sources ($m^*\ll m$) \cite{BW-KS:17, MG:17}. In fact, we find the source locations and the associated input parameters using an $\ell_1$-norm regularized optimization problem. In several inverse problems with sparsity constraints---finding unknown sparse parameters using a few measurements---$\ell_1$-norm regularization has shown to accurately recover the sparsity pattern of the unknown parameter than the standard  $\ell_2$-norm regularization \cite{JAT-SJW:10}. Sparsity has been long recognized in power system applications, especially in state and topology estimation problems; however, sparsity methods are less explored for localizing FO sources \cite{TH-NMF-PRK-LX:20}. 
	
	%Capitalizing on the sparsity and oscillatory nature of inputs injected by FO sources, 
%	It is possible to infer locations using either time-domain or frequency-domain measurements; see \cite{BW-KS:17, MG:17}, and references therein.
	{\color{black} By working in the frequency-domain, we encode sparsity of the number of locations and the number of sinusoids jointly in a single unknown vector $\mbf{U}_K$, where $\mbf{U}_K$ satisfies $\mbf{Y}_K=\mbf{G}_K\mbf{U}_K$ (see Section II for more details). This kind of joint sparsity might not be possible in the time-domain, especially, for arbitrary forced inputs \cite{RA-LS-OK:21}.} 
	
	We summarize our FO location recovery method below: 
	
	\begin{enumerate}
		{\item Using the discrete time Fourier transform (DTFT), we obtain the complex-valued linear model $\mbf{Y}_K=\mbf{G}_K\mbf{U}_K$ that relates measurements and the unknown oscillatory inputs at frequencies at which these inputs oscillate. Here, $K$ is the number of frequencies at all locations. 
			\item We then use the fast Fourier transform (FFT) to determine these frequencies by thresholding the spectrum of measurements collected over a finite time interval.}
		\item For the model in step 1, we solve an $\ell_1$ regularized least squares (henceforth, LASSO) problem in the complex domain via coordinated descent method \cite{JHF-TH-RT:10} to infer: (i) the number of true locations and the number of sinusoids at any given location, and (ii) the frequency, phase, and amplitude of each sinusoid. 
	\end{enumerate}
	Our recovery algorithm is simple as it requires thresholding the spectrum and solving a convex optimization problem. We validate its performance on two benchmark systems. 
	%{\color{blue}  . We do so using sparse learning using an $\ell_1$ regularizer. %-optimization method, which has found applications ranging from image and signal processing to bio-engineering. 
	
	%Given its simplicity and efficiency, we hope that our approach could be of interest to researchers and practitioners alike.
	%In practice, determining input frequencies by thresholding the empirical spectrum of measurements is well established. Building on this technique, we develop a recovery algorithm that infers input locations and characterstics of the input.

	%Overall our findings suggest {\color{red} What?} \ramargin{one line summary of our findings}
	
	%we use an  method to find  using the measurements and system model.
	% and only recently in forced oscillations  
	
	% and several methods are proposed to estimate state and topology  
	
	% is a key property that appear in
	
	% Importantly, these methods are shown to have remarkable performance (both theoretically and numerically) if the underlying unknown parameter vector is sparse (several components of the vector are zero) \cite{JHF-TH-RT:10}. A few of these methods have been applied to state and topology estimation in power systems as well .
	
	\smallskip
	\textit{Related literature}:  In \cite{NZ-MG-SA:17, ML-SL-DG-DW:20},  sources were identified using spectral properties of the transfer functions between sources and measurements. Along these lines, in \cite{SR-WJ-NN-BL:21}, sources are localized using ratios between signal content at the harmonics and the fundamental frequency. By representing generators' frequency response as an effective admittance matrix and oscillations as current injections, \cite{SCC-PV-KT:18} identifies sources by comparing predicted against measured current spectrum, and \cite{SCC-PV-KT:19} uses a Bayesian approach. Finally, in \cite{TH-NMF-PRK-LX:20}, a sparse plus low rank decomposition of Hankel measurement matrices has been suggested to localize the sources. 
	
	However, unlike our method, algorithms in the above studies localize sources (mostly single) but do not jointly estimate the input. Further, with the exception of works in \cite{TH-NMF-PRK-LX:20} and \cite{SCC-PV-KT:19}, sparsity has not been explored in a systematic way. {The work closest to ours is \cite{SCC-PV-KT:19}, which considers an $\ell_1$-norm regularized optimization framework in a Bayesian setting. However, \cite{SCC-PV-KT:19} models FOs as current injections and generators as frequency response functions. Instead, we model FOs as  exogenous inputs, which is applicable to any dynamical system, not just power systems. Moreover, we consider practical aspects, such as spectral leakage in the frequency domain.}

	%. These methods can be broadly classified as (i) data-driven methods, which rely only on the measurements and their features; (ii) model-based methods, which purely rely on the physics of the power system dynamics; and (iii) hybrid methods, which combine the advantages of the above approaches.
	
	%\textit{Organization}: The rest of the paper is organized as follows. Section II introduces the power system dynamics and its sampled transfer function model. Section III posits the localization problem as a convex optimization problem and solves it using the coordinate descent method. Finally, in Section VI, we validate the performance of the proposed method on several test cases and conclude the paper in Section V. 
	
	\section{System dynamics under forced oscillations}\label{sec: prelims}
	Consider the following linearized multi-machine dynamics excited by external unknown inputs in state-space form: 
	\begin{align}\label{eq: CT system}
		\begin{split}
			\dot{\bfx}(t)=\bfA\bfx(t)+\bfB\bfu(t),\quad \mbf{y}(t)=\mbf{C}\mbf{x}(t),\quad \forall t\geq 0.
		\end{split}
	\end{align}
	The state vector $\bfx(t)\in \mathbb{R}^{n}$ includes machine internal dynamics and variables associated with closed-loop controllers. We model the input $\bfu(t)\in \mathbb{R}^{m}$ as the forced oscillation vector:
	\begin{align}\label{eq: input sines}
		\mbf{u}(t)\triangleq \begin{bmatrix}
			u_1(t)\\
			u_2(t)\\
			\vdots \\
			u_m(t)\end{bmatrix}=\begin{bmatrix}
			\sum_{l=1}^{M_1}a_{1,l}\sin(\omega_{1,l}t+\phi_{1,l})\\
			\sum_{l=1}^{M_2}a_{2,l}\sin(\omega_{2,l}t+\phi_{2,l})\\
			\vdots \\
			\sum_{l=1}^{M_m}a_{m,l}\sin(\omega_{m,l}t+\phi_{m,l})
		\end{bmatrix}
	\end{align}
	where $a_{r,l}\geq 0$, $\omega_{r,l}=2\pi f_{r,l}\geq0 $, and $\phi_{r,l}$ are the amplitude, angular frequency, and phase of the $l^{th}$ sinusoid term at the $r^{th}$ location, and $r\in \{1,\ldots,m\}$. Finally, $\mbf{y}(t)\in \mathbb{R}^p$ is the measurement obtained from $p$ sensors. {For simplicity, we build the theory using noise-free measurements, and later consider measurement noise in simulations.}
	
	By modeling forced input as a sum of multiple weighted sinusoids, we allow for different oscillatory waveforms. Thus, our approach generalizes the approaches recovering a single sinusoidal input, which are well studied in the literature \cite{BW-KS:17, MG:17}. Let $\mbf{a}_r(t)=[a_{r,1}\sin(\omega_{r,1}t+\phi_{r,1})\,\cdots\, a_{r,M_r}\sin(\omega_{r,M_r}t+\phi_{r,M_r})]$ be the vector of sinusoids at the $r^{th}$ location. We make the following assumption on $\mbf{u}(t)$ in \eqref{eq: CT system} and $\mbf{a}_r(t)$. 
	
	\begin{assumption}\label{assump: sparsity} Both $\mbf{u}(t)$ in \eqref{eq: CT system} and $\mbf{a}_r(t)$ defined above are sparse; that is, $\|\mbf{u}(t)\|_0\ll m$ and $\|\mbf{a}_r(t)\|_0\ll M_r$, where $\|\mbf{z}\|_0$ counts the number of non-zero entries in the vector $\mbf{z}$.
	\end{assumption}
	
	As PMUs record measurements at discrete time instants, we sample \eqref{eq: CT system} with the sampling period $T$ (e.g., $F\!=\!1/T\!=\!30$--$60$ Hz for PMUs). Let $\mbf{A}_d=\exp(\mbf{A}T)$ and $\mbf{B}_d=\int_{0}^T\exp(\mbf{A}(T-s))ds\,\mbf{B}$. Let $k=0,1,\ldots$ and define $\mbf{x}[k]\triangleq \mbf{x}(kT)$, $\mbf{u}[k]\triangleq \mbf{u}(kT)$, and $\mbf{y}[k]\triangleq \mbf{y}(kT)$. Suppose that $\mbf{u}(t)$ is a piecewise constant\footnote{Frequencies of real FOs are smaller than the sampling frequency of PMUs.} during $kT\leq t \leq (k+1)T$. Then
	\begin{align}\label{eq: DT system}
		\begin{split}
			\hspace{-2.5mm}   \bfx[k+1]\!=\!\bfA_d\bfx[k]+\bfB_d\bfu[k],\quad \bfy[k]\!=\!\bfC\bfx[k]
		\end{split}
	\end{align}
	describe the discrete-time sampled model of \eqref{eq: CT system}. As our focus is on oscillations triggered by inputs, we assume $\mbf{x}[0]=\mbf{0}$.  
	
	Define the matrix-valued transfer function $\mbf{H}[z]\!=\!\mbf{C}(z\mbf{I}-\mbf{A}_d)^{-1}\bfB_d\in \mathbb{C}^{p\times m}$, where $z\in \mathbb{C}$. Then, from the standard linear system analysis, we conclude that $\mbf{Y}[z]\!=\!\mbf{H}[z]\mbf{U}[z]$, where $\mbf{Y}[z]\in \mathbb{C}^{p\times 1}$ and $\mbf{U}[z]\in \mathbb{C}^{m\times 1}$ are the $Z$-transforms of $\mbf{y}[k]$ and $\mbf{u}[k]$. Finally, at $z=\exp(j\Omega)$, where $\Omega\in (0, 2\pi)$ and $j^2=-1$, we have the DTFT representation of \eqref{eq: DT system}: 
	\begin{align}\label{eq: z-transfer function}
		\mbf{Y}[e^{j\Omega}]=\mbf{H}[e^{j\Omega}]\mbf{U}[e^{j\Omega}].
	\end{align}
	
	The benefit of working in the Fourier (or frequency) domain is that $\mbf{U}[e^{j\Omega}]$ can be expressed as a sum of weighted Dirac delta functions. As a result, $\mbf{Y}[e^{j\Omega}]$ can be expanded in terms of basis functions that encode both the source location and frequency of the input sinuosoids. Using this observation, we later show that the source localization problem can be cast as a solution to a simple $\ell_1$ regularized least-squares problem. 
	
	Since $\mbf{u}(t)$ in \eqref{eq: input sines} is a sum of sinusoids, it follows that 
	\begin{align}\label{eq: input delta functions}
		\hspace{-2.2mm}    \mbf{U}[e^{j\Omega}]\!=\!j\pi \begin{bmatrix}
			\sum_{l=1}^{M_1}a_{1,l}\,e^{j\phi_{1,l}}[\delta(\Omega\!+\!\wtilde{\omega}_{1,l})\!-\!\delta(\Omega\!-\!\wtilde{\omega}_{1,l})]\\
			\sum_{l=1}^{M_2}a_{2,l}\,e^{j\phi_{2,l}}[\delta(\Omega\!+\!\wtilde{\omega}_{2,l})\!-\!\delta(\Omega\!-\!\wtilde{\omega}_{2,l})]\\
			\vdots \\
			\sum_{l=1}^{M_m}a_{m,l}\,e^{j\phi_{m,l}}[\delta(\Omega\!+\!\wtilde{\omega}_{m,l})\!-\!\delta(\Omega\!-\!\wtilde{\omega}_{m,l})] 
		\end{bmatrix}
	\end{align}
	where $\delta(x)$ is the Dirac delta function and $\wtilde{\omega}_{k,l}=\omega_{k,l}T$. Note that $\mbf{U}[e^{j\Omega}]$ is non-zero only for $\Omega \in \{\pm \wtilde{\omega}_{1,1},\ldots, \pm \wtilde{\omega}_{m,M_m}\}$. Suppose that for $\Omega$ in this set, $\mbf{H}[e^{j\Omega}]\ne \mbf{0}$. Then, 
	from \eqref{eq: z-transfer function}, we have $\mbf{Y}[e^{j\Omega}]\ne \mbf{0}$. 
	If this is not the case, we cannot recover the sinusoid oscillating with $\Omega$ using measurements. 
	Finally, we have $\mbf{Y}[e^{j\Omega}]=\mbf{Y}^*[e^{-j\Omega}]$. Thus, we focus only on $\wtilde{\omega}_{k,l}\geq 0$. 
	
	% These frequencies are obtained by checking for which values of $\Omega \in (0,2\pi)$, $\mbf{Y}[e^{j\Omega}]\ne 0$. We comment more on this when we study discrete Fourier transform (DFT) using finite measurements. 

\begin{figure*}[t]
		\centering
		\includegraphics[width=1.0\linewidth]{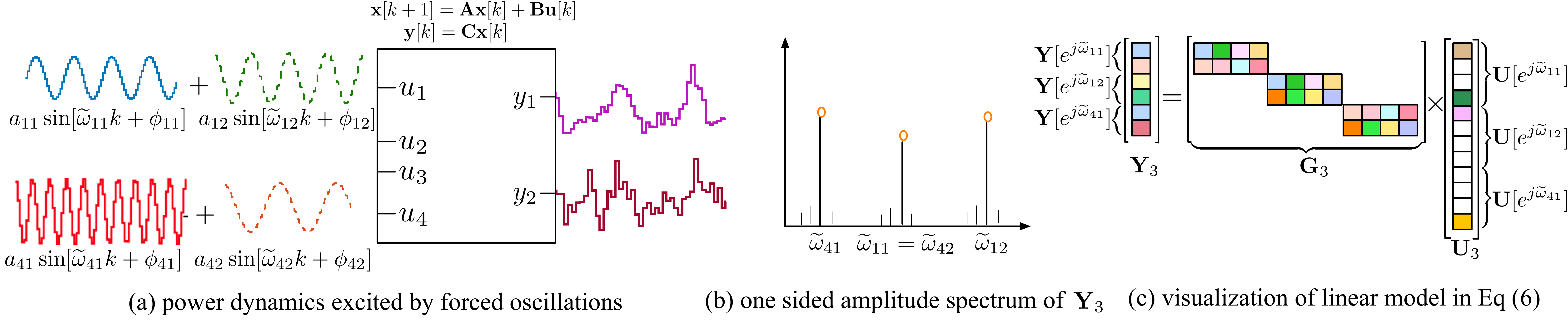} 
		\caption{\small (a) shows the discrete-time power system model with $m=4$ possible  locations but only locations $1$ and $4$ are injecting oscillations. Each of these inputs is a sum of two sinusoids with different parameters except for ${\widetilde{\omega}}_{11}={\widetilde{\omega}}_{42}$. In Fig.~(b), input frequencies  are found by computing the spectrum of the two sensory measurements. Since ${\widetilde{\omega}}_{11}={\widetilde{\omega}}_{42}$, we see only three dominant peaks in the spectrum, i.e., $K=3$, and the smaller peaks are due to spectral leakage. In Fig.~(c), we visualize linear model in \eqref{eq: block vector matrix}. For any $\widetilde{\omega}_{k,l}$, the four entries (square blocks) in $\mbf{U}[e^{j\widetilde{\omega}_{k,l}}]$ correspond to four possible locations. Locations having sinusoid oscillating with $\widetilde{\omega}_{k,l}$ are highlighted in color. The top input $\mbf{U}[e^{j\widetilde{\omega}_{1,1}}]$ has two non-zero entries (highlighted in brown and green) indicating that locations $1$ and $4$ have a sinusoid with $\widetilde{\omega}_{1,1}=\widetilde{\omega}_{4,2}$.}\label{fig: conceptual figure}
	\end{figure*}

	With a slight abuse of notation, let $\{\wtilde{\omega}_1,\ldots, \wtilde{\omega}_K\}$ be the set of frequencies where $\mbf{Y}[e^{j\wtilde{\omega}_l}]=\mbf{H}[e^{j\wtilde{\omega}_l}]\mbf{U}[e^{j\wtilde{\omega}_l}] \ne \bfzeros$, where both $\wtilde{\omega}_l$ and $K$ are unknown. Consider the following model:
	\begin{align}\label{eq: block vector matrix}
		\underbrace{\begin{bmatrix}
				\mbf{Y}[e^{j\wtilde{\omega}_1}]\\
				\mbf{Y}[e^{j\wtilde{\omega}_2}]\\
				\vdots \\
				\mbf{Y}[e^{j\wtilde{\omega}_K}]
		\end{bmatrix}}_{\triangleq \mbf{Y}_K}\!=\!
		\underbrace{\begingroup % keep the change local
			\setlength\arraycolsep{2pt}\begin{bmatrix}
				\mbf{H}[e^{j\wtilde{\omega}_1}] & \\
				& \mbf{H}[e^{j\wtilde{\omega}_2}]\\
				& & \ddots \\
				& & & \mbf{H}[e^{j\wtilde{\omega}_K}]
			\end{bmatrix}\endgroup}_{\triangleq \mbf{H}_K}
		\underbrace{\begin{bmatrix}
				\mbf{U}[e^{j\wtilde{\omega}_1}]\\
				\mbf{U}[e^{j\wtilde{\omega}_2}]\\
				\vdots \\
				\mbf{U}[e^{j\wtilde{\omega}_K}]
		\end{bmatrix}}_{\triangleq \mbf{U}_K}
	\end{align}
	where $\mbf{Y}_K \in \mathbb{C}^{pK\times 1}$, $\mbf{H}_K\in \mathbb{C}^{pK\times mK}$, and $\mbf{U}_K \in \mathbb{C}^{mK\times 1}$.
	In light of Assumption \ref{assump: sparsity}, note that $K\ll [M_1+\ldots+M_m]$. Further, the non-zero components of $\mbf{U}[e^{j\wtilde{\omega}_l}]\in \mathbb{C}^{m}$ correspond to locations with the inputs containing sinusoids of frequency $\wtilde{\omega}_l$. Thus, from the sparsity pattern of $\mbf{U}_K$ alone we can determine source locations and their sinusoidal frequencies (see Fig.~\ref{fig: conceptual figure}). The values of $\mbf{U}_K$ provide phase and amplitude information. 
	
	The model in \eqref{eq: block vector matrix} captures the relationship between measurements and the input locations at multiple frequencies; thus, for a sinusoid input, i.e., $\mbf{U}[e^{j\wtilde{\omega}_l}]$
	is a delta function, $\mbf{Y}[e^{j\wtilde{\omega}_l}]$ is infinite for any $l$. This is a natural aspect of taking DTFTs of infinite time-length sinusoids. %This behavior is expected as we assume $\mbf{U}[e^{j\omega}]$ is the DTFT of the infinite-length oscillatory inputs $\mbf{u}[k]$. 
	However, in practice, we only use measurements for a finite time interval and use FFTs to compute $\mbf{Y}[e^{j\wtilde{\omega}_l}]$, which is consequently finite. % (see Section III for more details). 

	\section{Complex-LASSO for Source Localization}
	
	{\color{black} In practice, the total number of sensors (p) could be less than the possible number of sources (m). For example, PMUs can measure only bus level quantities but not the internal signals of control devices. Thus, model in \eqref{eq: block vector matrix} is an under-determined system and we cannot obtain $\mbf{U}_K$ by means of ordinary least squares.} As a result, 
	for the model in \eqref{eq: block vector matrix}, we consider the $\ell_1$-regularized optimization problem to both localize the sources and identify the frequencies of the sinusoids at each source:
	\begin{align}\label{eq: group LASSO opt}
		\what{\mbf{U}}_K\!=\!\argmin_{\substack{\bfU_K \in \mathbb{C}^{mK\times 1} }}\left\lbrace\frac{1}{2}\norm{\mbf{Y}_K-\mbf{H}_K\bfU_K}_2^2\!+\!\lambda \|\mbf{U}_K\|_1\right\rbrace
	\end{align}
	where $\lambda\geq 0$ is the tuning parameter. For the vector  $\mathbf{z}$, the $\ell_1$-norm is $\|\mathbf{z}\|_1=\sum_i |z_i|$, where $|z_i|=\sqrt{\text{Re}(z_i)^2+\text{Im}(z_i)^2}$. The regularization term $\|\mbf{U}_K\|_1$ 
	%$\|\mbf{U}_K\|_1=\sum_{l=1}^{mK}|\mbf{U}_K(i)|$  (with $\mbf{U}_K(i)$ being the $i^{th}$ scalar entry of $\mbf{U}_K$) is the $\ell_1$-norm, which Note that $\|\mbf{U}_K\|_1=\sum_{l=1}^{K}\|\mbf{U}_K[e^{j\wtilde{\omega}_l}]\|_1$.
	promotes sparsity in $\what{\mbf{U}}_K$.  We henceforth refer to the problem in \eqref{eq: group LASSO opt} as the complex-LASSO. 
	
	Akin to $\mbf{U}_K$ in \eqref{eq: block vector matrix}, define
	$\what{\mbf{U}}_K^\transpose=[\what{\mbf{U}}[e^{j\wtilde{\omega}_1}]^\transpose\ldots \what{\mbf{U}}[e^{j\wtilde{\omega}_K}]^\transpose]$. The block diagonal form of $\mbf{H}_K$ and additive property of the $\ell_1$-norm allow us to compute $\what{\mbf{U}}[e^{j\wtilde{\omega}_l}]$ in a distributed fashion:
	\begin{align*}
		\what{\mbf{U}}[e^{j\wtilde{\omega}_l}]\!=\!\argmin_{\substack{\bfU \in \mathbb{C}^{m\times 1} }}\left\lbrace\frac{1}{2}\norm{\mbf{Y}[e^{j\wtilde{\omega}_l}]-\mbf{H}[e^{j\wtilde{\omega}_l}]\bfU}_2^2\!+\!\lambda \|\mbf{U}\|_1\right\rbrace. 
	\end{align*}
	The above optimization problem is computationally convenient and extremely useful to quickly determine the sources injecting oscillations with a particular frequency of interest. 
	
%	Theoretical performance of the estimator in \eqref{eq: group LASSO opt}, in the context of linear regression models, has been recently established in \cite{RMW:20}. Importantly, \cite{RMW:20} notes that $\what{\mbf{U}}_K$ accurately estimates $\mbf{U}_K$ and its sparsity pattern, provided the input-to-noise ratio is not too small. 
	To solve \eqref{eq: group LASSO opt}, we need $\mbf{H}_K$, $\mbf{Y}_K$, and $K$. The matrix $\mbf{H}_K$ is computed using the power system matrices $(\mbf{A}_d, \mbf{B}_d, \mbf{C})$. If not available, one can use empirically determined transfer functions. To compute $K$ and $\mbf{Y}_K$, we obtain the vector-valued $N$-point DFT\footnote{In simulations, for computational speedup, we use the FFT.} of $\mbf{y}[L],\ldots,\mbf{y}[N-1+L]$ using 
	\begin{align}\label{eq: FFT}
		\wtilde{\mbf{Y}}[q]\triangleq \frac{2}{N}\sum_{k=L}^{N-1+L}\mbf{y}[k]e^{-j\frac{2\pi l k}{N}} \quad (q=0,1,\ldots,N-1)
	\end{align}
	where we choose $L\gg 0$ such that $\mbf{y}[k]$, for all $k \geq L$, is in steady state. For any $q$, recall that $\|\wtilde{\mbf{Y}}[q]\|_\infty=\max_{i}|\wtilde{\mbf{Y}}^{(i)}[q]|$, where $\wtilde{\mbf{Y}}^{(i)}[q]$ is the $i^{th}$ scalar in $\wtilde{\mbf{Y}}^{(i)}[q]\in \mathbb{C}^{p}$. 
	
	The input (angular) frequencies can be determined using $\widetilde{\omega}_l=2\pi l/N$, where $l$ satisfies $\|\wtilde{\mbf{Y}}[l]\|_\infty\ne 0$. However, due to the sensor noise and spectral leakage, $\|\wtilde{\mbf{Y}}[l]\|_\infty$ could be non-zero even when $\widetilde{\omega}_l$ is not the true input frequency. The noise can be attenuated by filtering the measurements. Instead, we reduce spectral leakage by multiplying measurements with the Hamming window \cite{PS-RM:97}. For these processed measurements, let $S=\{l: \|\wtilde{\mbf{Y}}[l]\|_\infty> \tau\}$, where $\tau >0$ is the user-defined threshold. Then $K=|S|$, where $|\cdot|$ is the cardinality of the set $S$, gives us the total number of input frequencies. Finally, we replace $\mbf{Y}[e^{j\widetilde{\omega}_l}]$ in $\mbf{Y}_K$ (given by \eqref{eq: block vector matrix}) with $\wtilde{\mbf{Y}}[l]$, and evaluate the corresponding $\mbf{H}[e^{j\widetilde{\omega}_l}]$ at $\widetilde{\omega}_l=2\pi l/N$, where $l \in S$. 
	
	% Without of loss of generality, let $S=\{1,2,\ldots,L\}$ and obtain $\mbf{Y}_K$ in 
	% \eqref{eq: block vector matrix} by setting 
	% $\mbf{Y}_K^\transpose=[\wtilde{\mbf{Y}}^\transpose[l],\ldots,\wtilde{\mbf{Y}}^\transpose[K]]$. \ramargin{change this last line}
	% $\{\mbf{y}[k]\}_{k=L}^{N-1}$.
	
	% {\color{blue} Instead, we reduce spectral leakage by multiplying (element-wise) every component series of $\{\mbf{y}[k]\}_{k=L}^{N-1}$ with the Hamming window: $0.54-0.46\cos(2\pi n/N)$, where $0\leq n\leq N-1$. We then compute the spectrum of the resultant $\{\mbf{y}[k]\}_{k=L}^{N-1}$. This operation reduces spectral leakage by smoothing discontinuities at both end points of $\{\mbf{y}[k]\}_{k=L}^{N-1}$.}

	% Let  and define $S\triangleq \{q: \|\wtilde{\mbf{Y}}[q]\|_\infty\geq \tau\}$, where  $\|\wtilde{\mbf{Y}}[q]\|_\infty$ is the maximum of (in magnitude) entries in $\wtilde{\mbf{Y}}[q]\in \mathbb{C}^{p}$. Then, $K=|S|$, that is, the number of entries in $S$. 
	
	With $\mbf{H}_K$ and $\mbf{Y}_K$ at our disposal, we now can solve \eqref{eq: group LASSO opt} using coordinate descent method \cite{JHF-TH-RT:10} in the complex domain. We summarize our source recovery method in Algorithm \ref{alg: complex LASSO}, which recovers locations, and input parameters in discrete-time domain. We obtain the continuous-time frequency as $f_l= \wtilde{\omega}_{l}/(2\pi T)$. Instead, we recover amplitudes by $|\what{\mbf{U}}_K|/\pi$, where $|\cdot|$ is the complex magnitude, applied for entry in $\what{\mbf{U}}_K$. 
	
		\vspace{-2.0mm}
	%{\color{red} NIMA state} The amplitude and the phase of the input parameters are estimated by computing the magnitude and phase of the complex numbers in $\what{\mbf{U}}_K/j\pi$ (as shown in \eqref{eq: input delta functions}). 
	% \footnote{Because these are computed via $\mathcal{Z}$-transform by assuming the availability of infinite number of measurements.}
	\SetKwInput{KwSa}{Step 1}  
	\SetKwInput{KwSaa}{Step 2}  
	\SetKwInput{KwSb}{Step 3}
	\SetKwInput{KwSc}{Step 4}
	\SetKwInput{KwInp}{Input} 
	\SetKwInput{KwReturn}{Return} 
	\begin{algorithm}
		\DontPrintSemicolon
		\SetAlgoLined
		\KwInp{$\tau>0$, $\lambda\geq 0$, $\{\mbf{Y}[k]\}_{k=L}^{N-1-L}$, and $(\mbf{A}_d,\mbf{B}_d,\mbf{C})$.}
		\KwSa{Compute $\widetilde{\mbf{Y}}[q]$ using \eqref{eq: FFT} and process them using the Hamming window.}
		\KwSaa{Set $S=\{l: \|\wtilde{\mbf{Y}}[l]\|_\infty> \tau\}$ and $K=|S|$.}
		\KwSb{Let $\mbf{Y}_K=[\wtilde{\mbf{Y}}[l_1]\ldots \wtilde{\mbf{Y}}[l_K]]$. Evaluate $\mbf{H}[e^{j\wtilde{\omega}_{l_i}}]$ in \eqref{eq: block vector matrix} using $\wtilde{\omega}_{l_i}=2\pi l_i/N$, and $l_i \in S$} 
		\KwSc{Solve complex-LASSO problem in \eqref{eq: group LASSO opt} using the coordinate descent method \cite{JHF-TH-RT:10}.}
		\KwReturn{$\wtilde{\omega}_l$; $\what{\mbf{U}}[e^{j\wtilde{\omega}_l}]$; and source locations: indices of non-zero entries in $\what{\mbf{U}}[e^{j\wtilde{\omega}_l}]$, $l=\{1,\ldots,K\}$.} 
		\caption{Source localization via complex-LASSO}\label{alg: complex LASSO}
	\end{algorithm}
	% 	{\color{red} Finally, we recover continuous-time (angular) frequency using $\widetilde{\omega}_l={\omega}_lT$, where $T$ is the sampling period.}
	
	\vspace{-5.0mm}
	\section{Simulation Results}
	We apply Algorithm \ref{alg: complex LASSO} to recover FO locations and oscillatory input parameters in two benchmark power systems. We add Gaussian noise with the SNR of 10 dB to the measurable quantities (see below). We let $1/T=F=30$ and $N=600$ .  
	
	{\bf Tuning parameter ($\lambda$) selection}: For $\lambda=0$, the solution to \eqref{eq: group LASSO opt} is given by the least squares solution. For $\lambda\geq \lambda_{\max}\triangleq \|\mbf{Y}_K^T \mbf{H}_K \|_\infty$, $\what{\mbf{U}}_K=\mbf{0}$  (a fully sparse vector) \cite[Section 2.5]{JHF-TH-RT:10}. Thus, we set $\lambda = \alpha \lambda_{\text{max}}$ and choose $\alpha\in [0,1]$ by performing a sensitivity analysis with respect to the true positive rate (TPR)---ratio of the number of correctly identified non-zero entries in $\what{\mbf{U}}_K$ to the number of non-zero entries in $\mbf{U}_K$---and the false positive rate (FPR)---ratio of the number of incorrectly identified non-zero entries in $\what{\mbf{U}}_K$ to the number of non-zero entries in $\mbf{U}_K$.
%	
%	
%	
%	and false positive rate (FPR): 
%	\begin{equation}
%		\begin{split}
%			\text{TPR} & = \frac{\text{number of correct non-zero elements in  $\what{\mbf{U}}_K$}}{\text{number of non-zero elements in} {\mbf{U}}^*_K }\\
%			\text{FPR} & = \frac{\text{number of incorrect non-zero elements in $\what{\mbf{U}}_K$}}{\text{number of zero elements in} {\mbf{U}}^*_K }.
%		\end{split}
%	\end{equation}
%	
	{\color{black} TPR and FPR depend on the noisy measurements and so does $\alpha$. Thus to get a reliable and trustworthy $\alpha$, we perform sensitivity analysis for 20 realizations of $\{y[k]\}_{k=L}^{N-1+L}$. We then pick an $\alpha$ that yields $\text{TPR}=1$ and $\text{FPR}=0$ for as many realizations as possible.}

	% we proceed as follows. We first generate  and compute $\alpha$

	% we compute TPR and FPR for  We then pick an $\alpha$ 

	% \textcolor{red}{ To investigate the impact of measurement noise on the performance of the proposed algorithm, the TPR and FPR values are calculated over 20 realization of noisy measurements by adding a Gaussian white noise to 
	% $y[k]$, and $k=L,\cdots, N-1+L$. }
	
	% We choose the tuning parameter $\lambda$ by performing sensitivity analysis $\lambda\geq \in [0,\lambda_\text{max}]$ as follows: first, we write 
	% \begin{remark}
	% The regularization parameter can be written as  where $\lambda_{\text{max}} = \|\mbf{Y}_K^T \mbf{H}_K \|_\infty$ and $0\le \alpha \le 1$. The regularization parameter $\lambda$ can be obtained by 
	% \end{remark}

	% \usepackage{multirow}
	\begin{table*}[]
		\begin{threeparttable}
			\centering
			\caption{True and estimated values of the amplitude, frequency and phase of the forced oscillations inputs}\label{Table: Inputs}
			\begin{tabular}{|c|c|ccccccccc|}
				\hline
				\multirow{3}{*}{\textbf{\begin{tabular}[c]{@{}c@{}}System \\ under \\ study\end{tabular}}} & \multirow{3}{*}{\textbf{\begin{tabular}[c]{@{}c@{}}FO\\ Location\end{tabular}}} & \multicolumn{9}{c|}{\textbf{Input Signal}}                                                                                                                                                                                                                                                                                                                                                                 \\ \cline{3-11} 
				&                                                                                 & \multicolumn{3}{c|}{\textbf{Amplitude (p.u.)}}                                                                                                          & \multicolumn{3}{c|}{\textbf{Frequency (Hz)}}                                                                                & \multicolumn{3}{c|}{\textbf{Phase (rad)}}                                                                                            \\ \cline{3-11} 
				&                                                                                 & \multicolumn{1}{c|}{\textbf{Parameter}}  & \multicolumn{1}{c|}{\textbf{True}} & \multicolumn{1}{c|}{\textbf{Estimate ($\sigma$)\tnote{*}}} & \multicolumn{1}{c|}{\textbf{Parameter}}  & \multicolumn{1}{c|}{\textbf{True}} & \multicolumn{1}{c|}{\textbf{Estimate}} & \multicolumn{1}{c|}{\textbf{Parameter}}     & \multicolumn{1}{c|}{\textbf{True}} & \textbf{Estimate ($\sigma$)\tnote{*}} \\ \hline
				\multirow{6}{*}{\textbf{\begin{tabular}[c]{@{}c@{}}68-bus system\\ (near PMUs)\end{tabular}}}         & \multirow{2}{*}{4}                                                              & \multicolumn{1}{c|}{\textbf{$a_{4,1}$}}  & \multicolumn{1}{c|}{$0.01$}        & \multicolumn{1}{c|}{$0.0126~(0.003)$}                             & \multicolumn{1}{c|}{\textbf{$f_{4,1}$}}  & \multicolumn{1}{c|}{$2$}           & \multicolumn{1}{c|}{$2$}               & \multicolumn{1}{c|}{\textbf{$\phi_{4,1}$}}  & \multicolumn{1}{c|}{$0.3$}         & $0.3302~(0.024)$                             \\ \cline{3-11} 
				&                                                                                 & \multicolumn{1}{c|}{\textbf{$a_{4,2}$}}  & \multicolumn{1}{c|}{$0.01$}        & \multicolumn{1}{c|}{$0.0081~(0.0018)$}                            & \multicolumn{1}{c|}{\textbf{$f_{4,2}$}}  & \multicolumn{1}{c|}{$2.5$}         & \multicolumn{1}{c|}{$2.5$}             & \multicolumn{1}{c|}{\textbf{$\phi_{4,2}$}}  & \multicolumn{1}{c|}{$0.4$}         & $0.4401~(0.012)$                             \\ \cline{2-11} 
				& \multirow{2}{*}{10}                                                             & \multicolumn{1}{c|}{\textbf{$a_{10,1}$}} & \multicolumn{1}{c|}{$0.01$}        & \multicolumn{1}{c|}{$0.0087~(0.005)$}                             & \multicolumn{1}{c|}{\textbf{$f_{10,1}$}} & \multicolumn{1}{c|}{$1.5$}         & \multicolumn{1}{c|}{$1.5$}             & \multicolumn{1}{c|}{\textbf{$\phi_{10,1}$}} & \multicolumn{1}{c|}{$0.1$}         & $0.1326~(0.01)$                              \\ \cline{3-11} 
				&                                                                                 & \multicolumn{1}{c|}{\textbf{$a_{10,2}$}} & \multicolumn{1}{c|}{$0.01$}        & \multicolumn{1}{c|}{$0.0093~(0.006)$}                             & \multicolumn{1}{c|}{\textbf{$f_{10,2}$}} & \multicolumn{1}{c|}{$1$}           & \multicolumn{1}{c|}{$1$}               & \multicolumn{1}{c|}{\textbf{$\phi_{10,2}$}} & \multicolumn{1}{c|}{$0.3$}         & $0.2804~(0.024)$                             \\ \cline{2-11} 
				& \multirow{2}{*}{13}                                                             & \multicolumn{1}{c|}{\textbf{$a_{13,1}$}} & \multicolumn{1}{c|}{$0.02$}        & \multicolumn{1}{c|}{$0.0169~(0.004)$}                             & \multicolumn{1}{c|}{\textbf{$f_{13,1}$}} & \multicolumn{1}{c|}{$3.5$}         & \multicolumn{1}{c|}{$3.5$}             & \multicolumn{1}{c|}{\textbf{$\phi_{13,1}$}} & \multicolumn{1}{c|}{$0.1$}         & $0.0893~(0.007)$                             \\ \cline{3-11} 
				&                                                                                 & \multicolumn{1}{c|}{\textbf{$a_{13,2}$}} & \multicolumn{1}{c|}{$0.02$}        & \multicolumn{1}{c|}{$0.0170~(0.002)$}                             & \multicolumn{1}{c|}{\textbf{$f_{13,2}$}} & \multicolumn{1}{c|}{$0.8$}         & \multicolumn{1}{c|}{$0.8$}             & \multicolumn{1}{c|}{\textbf{$\phi_{13,2}$}} & \multicolumn{1}{c|}{$0.2$}         & $0.1771~(0.0052)$                            \\ \hline
				\multirow{6}{*}{\textbf{\begin{tabular}[c]{@{}c@{}}WECC-179\\  bus system\end{tabular}}}   & \multirow{2}{*}{5}                                                              & \multicolumn{1}{c|}{\textbf{$a_{5,1}$}}  & \multicolumn{1}{c|}{$0.02$}        & \multicolumn{1}{c|}{$0.0244~(0.002)$}                             & \multicolumn{1}{c|}{\textbf{$f_{5,1}$}}  & \multicolumn{1}{c|}{$1$}           & \multicolumn{1}{c|}{$1$}               & \multicolumn{1}{c|}{\textbf{$\phi_{5,1}$}}  & \multicolumn{1}{c|}{$0.3$}         & $0.2993~(0.0133)$                            \\ \cline{3-11} 
				&                                                                                 & \multicolumn{1}{c|}{\textbf{$a_{5,2}$}}  & \multicolumn{1}{c|}{$0.01$}        & \multicolumn{1}{c|}{$0.0117~(0.005)$}                             & \multicolumn{1}{c|}{\textbf{$f_{5,2}$}}  & \multicolumn{1}{c|}{$0.8$}         & \multicolumn{1}{c|}{$0.8$}             & \multicolumn{1}{c|}{\textbf{$\phi_{5,2}$}}  & \multicolumn{1}{c|}{$0.4$}         & $0.3405~(0.0305)$                            \\ \cline{2-11} 
				& \multirow{2}{*}{14}                                                             & \multicolumn{1}{c|}{\textbf{$a_{14,1}$}} & \multicolumn{1}{c|}{$0.03$}        & \multicolumn{1}{c|}{$0.0214~(0.008)$}                             & \multicolumn{1}{c|}{\textbf{$f_{14,1}$}} & \multicolumn{1}{c|}{$0.7$}         & \multicolumn{1}{c|}{$0.7$}             & \multicolumn{1}{c|}{\textbf{$\phi_{14,1}$}} & \multicolumn{1}{c|}{$0.2$}         & $0.2134~(0.0068)$                            \\ \cline{3-11} 
				&                                                                                 & \multicolumn{1}{c|}{\textbf{$a_{14,2}$}} & \multicolumn{1}{c|}{$0.02$}        & \multicolumn{1}{c|}{$0.0170~(0.001)$}                             & \multicolumn{1}{c|}{\textbf{$f_{14,2}$}} & \multicolumn{1}{c|}{$1.5$}         & \multicolumn{1}{c|}{$1.5$}             & \multicolumn{1}{c|}{\textbf{$\phi_{14,2}$}} & \multicolumn{1}{c|}{$0.3$}         & $0.2613~(0.0048)$                            \\ \cline{2-11} 
				& \multirow{2}{*}{27}                                                             & \multicolumn{1}{c|}{\textbf{$a_{27,1}$}} & \multicolumn{1}{c|}{$0.04$}        & \multicolumn{1}{c|}{$0.0350~(0.007)$}                             & \multicolumn{1}{c|}{\textbf{$f_{27,1}$}} & \multicolumn{1}{c|}{$2$}           & \multicolumn{1}{c|}{$2$}               & \multicolumn{1}{c|}{\textbf{$\phi_{27,1}$}} & \multicolumn{1}{c|}{$0.1$}         & $0.1076~(0.0098)$                            \\ \cline{3-11} 
				&                                                                                 & \multicolumn{1}{c|}{\textbf{$a_{27,2}$}} & \multicolumn{1}{c|}{$0.01$}        & \multicolumn{1}{c|}{$0.0094~(0.003)$}                             & \multicolumn{1}{c|}{\textbf{$f_{27,2}$}} & \multicolumn{1}{c|}{$1.2$}         & \multicolumn{1}{c|}{$1.2$}             & \multicolumn{1}{c|}{\textbf{$\phi_{27,2}$}} & \multicolumn{1}{c|}{$0.2$}         & $0.2567~(0.0457)$                            \\ \hline
			\end{tabular}
			\begin{tablenotes}\footnotesize
				\item[*] Standard deviation of estimated parameters over 20 realizations of noisy measurements.
			\end{tablenotes}
		\end{threeparttable}
	\end{table*}

	\begin{figure*}
	\centering
	\includegraphics[width=1.0\linewidth]{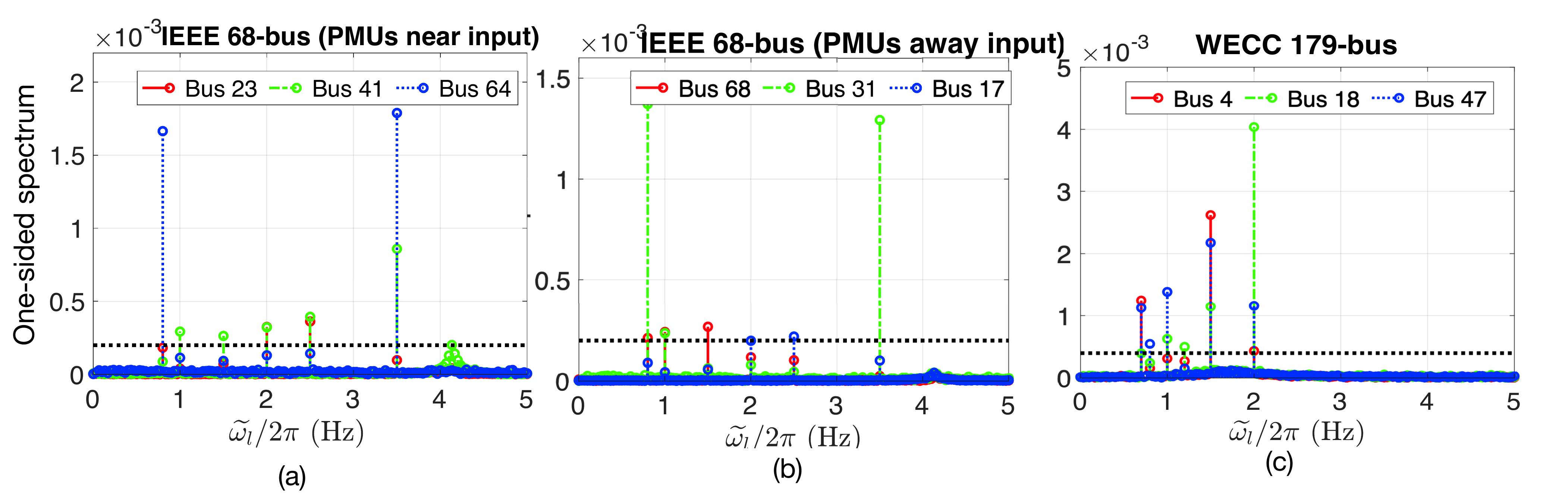} 
	\caption{\small \textcolor{black}{One-sided power spectrum of measurements for different case studies. In each plot, we overlay the spectrum associated with different measurement buses. The black dashed line represents the thresholding parameter $\tau$, which equals $0.2\times 10^{-3}$ in Fig.~(a) and (b), and $0.4 \times 10^{-3}$ in Fig.~(c). Theoretically, if inputs are observable, the spectrum of each PMU measurement should contain dominant peaks at all input frequencies. However, as evident in all plots, not even for one PMU, we can see dominant peaks at all input frequencies. This can be attributed to the measurement noise and spectral leakage. However, by considering all PMUs, we see sharp dominant peaks at all input frequencies. Interestingly, for IEEE-68 bus case, the magnitude of spectral peaks is smaller when PMUs are away from the true source locations. Thus it makes sense to consider measurements from nearby PMUs if we have prior coarse knowledge of source locations.}}\label{fig: power_spectrum}
\end{figure*}	
	
	\begin{figure*}
		\centering
		\includegraphics[width=1.0\linewidth]{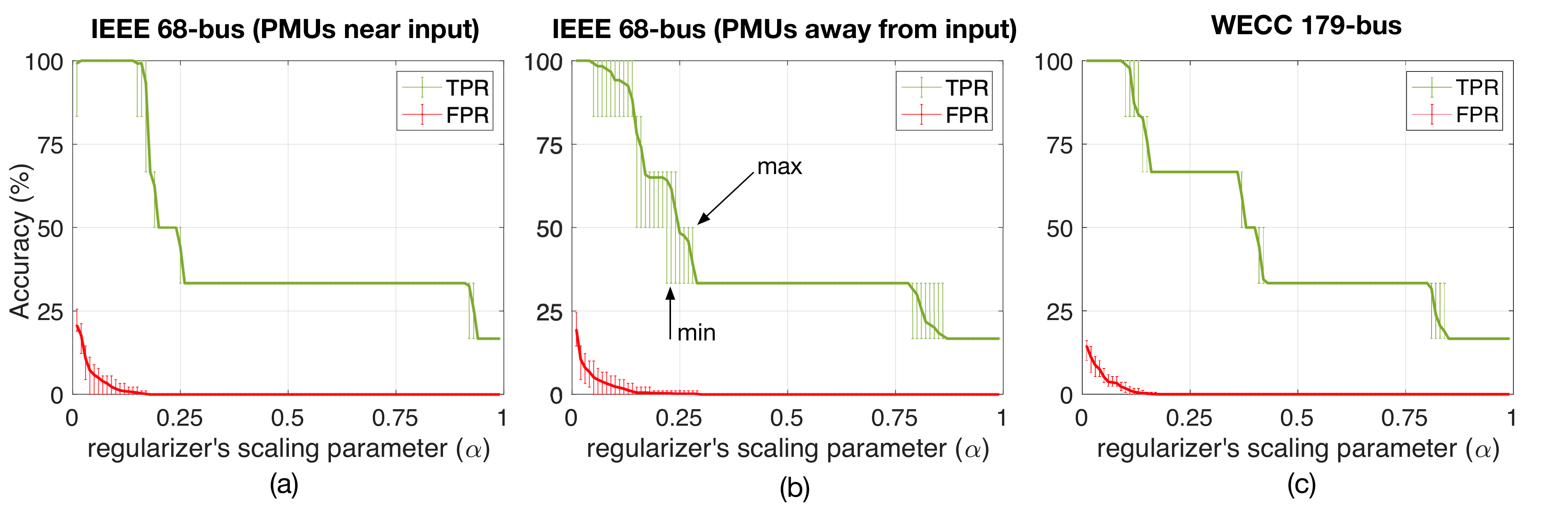} 
		\caption{\small Sensitivity analysis of the scaling parameter $\alpha$ in terms of TPR and FPR for different cases. 
			%For the 3 machine system, observe that we can identify the correct source at bus 1 only for a small range of values of $\alpha$ (around 0.05). This is because the input is not too sparse; that is, $m^*/m=1/3\approx 0.33$. 
			{The error bars in the figures illustrate the maximum and minimum of twenty values (as described in the main text) of TPR and FPR. Instead, the solid line is the average value}
			Interestingly, from panels (a) and (b), we note that the PMUs in the vicinity of the FO sources resulted in a wider range of $\alpha$ values that yield a better TPR ($100 \%$) and FPR ($0\%$). }
		\label{fig: TPR_FPR}
	\end{figure*}

	\subsection{Case 1: IEEE 68 bus, 16 machine system \cite{68B16M}}
	\vspace{-3.0mm}
	\begin{figure}[H]
		\centering
		\includegraphics[width=0.9\linewidth]{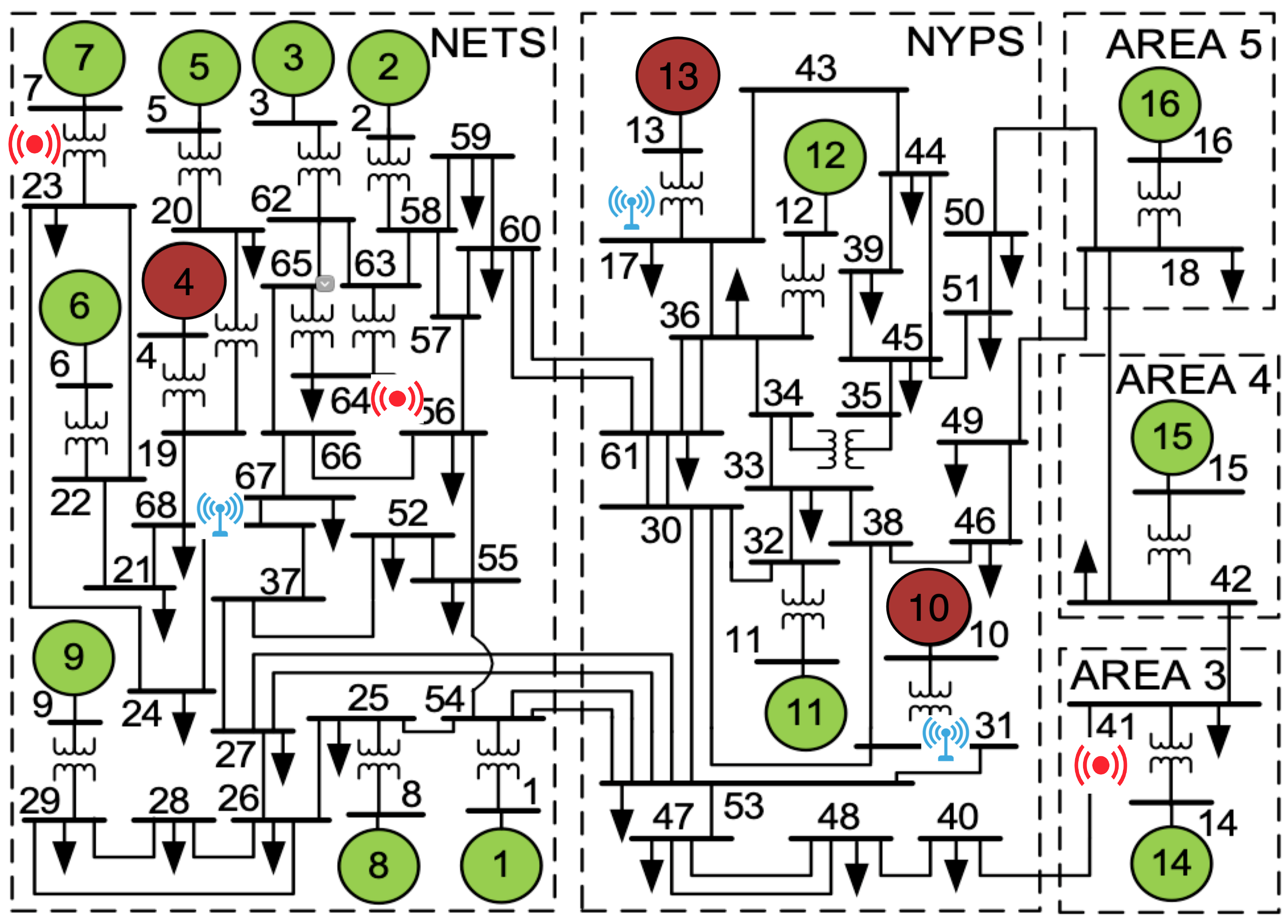} 
		\caption{\small IEEE NETS/NYPS 16 machine 68 bus system \cite{68B16M}. Among $m=16$ excitation control inputs (green circles), only $m^*=3$ locations (red colored bus no.s: $4$, $10$, $13$) are excited by FO inputs.} \label{fig: sixteen_bus}
	\end{figure}
	
	\vspace{-3.0mm}
	Each generator is represented by ten states: the rotor angle, angular frequency, damper winding flux leakages, and states corresponding to excitation systems. We obtain the state space matrices using the power system toolbox (PST) \cite{PST}. See Fig.~\ref{fig: sixteen_bus} and Table \ref{Table: Inputs} for the inputs' description. We consider scenarios where PMUs are near and far away from the sources. For the first, PMUs are at buses 68, 31, and 17 (highlighted with a blue sensor icon in Fig.~\ref{fig: sixteen_bus}). For the latter, we place PMUs at buses 23, 41, and 64 (highlighted with a red sensor icon in Fig.~\ref{fig: sixteen_bus}). For all realizations, we correctly determined the input frequencies by thresholding the power spectrum of the voltage magnitude measurements (near and far case). Fig. \ref{fig: power_spectrum} shows the spectrum of an arbitrary measurement realization. 

	The sensitivity analysis of the parameter $\alpha$ in terms of TPR and FPR for the two scenarios are shown in Fig. \ref{fig: TPR_FPR}~(a) and (b), respectively. Due to space limitations, we focus only on nearby PMUs scenario. For $0.08 \le \alpha \le 0.14$, we were able to accurately find true FO locations in most of the realizations. Finally, for $\alpha=0.14$, we report average and standard deviation of estimated input parameters in Table \ref{Table: Inputs}.

	\subsection{Case 2: Reduced WECC 179 bus 29 machine system \cite{TESTCASE}}
	%For further validation, we apply the proposed method on the WECC 179  bus system \cite{TESTCASE} in the presence of multiple FOs. 
	Each generator is modeled as a second order classical model and has two states: rotor angle and angular frequency. We use the small signal analysis tool (SSAT) to extract state space matrices of the WECC model described in \cite{TESTCASE}. We add FO input signals (parameters reported in Table \ref{Table: Inputs}) to the mechanical torque of three generators at buses, labeled 5, 14, and 27 in \cite[Fig.~1]{TESTCASE}. The PMU buses are 4, 18, and 47. Our measurements consist of the rotor angle of generators \cite{ML-SL-DG-DW:20}. 
	% and are able to measure the rotor angle of the generators. The input signals (as shown in Table \ref{Table: Inputs}) are added to the mechanical torque of three generators across the grid. 
	Fig.~\ref{fig: TPR_FPR}~(c) illustrates the sensitivity analysis of the parameter $\alpha$ in terms of TPR and FPR. Note that for $\alpha = 0.11$, we found the correct source locations. 
	%\begin{figure}[h!]
	%	\centering
	%	\includegraphics[scale = 0.5]{IEEE_PESGM2022/wecc179_basecase.jpg}
	%	\caption{Reduced WECC 179 bus 29 machine system \cite{68B16M}}\label{fig: WECC_SLD}
	%\end{figure}
	
	% \begin{figure}[h!]
	% 	\centering
	% 	\includegraphics[scale = 0.6,trim={4cm 8.5cm 4cm 8.5cm},clip]{IEEE_PESGM2022/179b_alpha.pdf} 
	% 	\caption{Sensitivity analysis of the parameter $\alpha$ in terms of TPR and FPR in identifying the source of the FOs in the WECC 179 bus 29 machine system.}\label{fig: 179b_alpha}
	% \end{figure}

	%\clearpage
	%\pagebreak 
	\section{Conclusions}
	This paper studies the effectiveness of complex-LASSO for (i) localizing sources of multiple forced oscillatory inputs and (ii) estimating the input parameters. We show that sparsity in the number of locations and sparsity in the number of sinusoids at a location are clearly manifested in the frequency domain. This observation led us to cast the localization problem as an $\ell_1$-regularized least squares problem, which we solve numerically via a complex-valued coordinate descent method. Our localization-estimation Algorithm \ref{alg: complex LASSO} is simple and has a potential to be integrated into real-time grid operations. For simplicity, we present our results assuming the knowledge of dynamic models; However, our approach is general and can even work with empirically determined transfer functions.
	
	\bibliographystyle{unsrt}
	\bibliography{BIB.bib}
\end{document}